%Paper: hep-th/9211136
%From: FUCITO@roma2.infn.it
%Date: Mon, 30 Nov 1992 12:17:57 +0100 (WET)

%Updated with effect from: 5 Sept 1991
\magnification=\magstep1
\headline={\ifnum\pageno=1\firstheadline\else
\ifodd\pageno\rightheadline \else\leftheadline\fi\fi}
\def\firstheadline{\hfil}
\def\rightheadline{\hfil}
\def\leftheadline{\hfil}
        \footline={\ifnum\pageno=1\firstfootline\else\otherfootline\fi}
\def\firstfootline{\rm\hss\folio\hss}
\def\otherfootline{\hfil}

\font\tenrm=cmr10
\font\tenit=cmti10
\font\elevenbf=cmbx10 scaled\magstep 1
 1
 1

\def\de{\partial}

\def\intx{{1\over 2\pi}\int d^2 x\ }
\def\kinx#1{\de_\mu #1\de^\mu #1}
\def\curv#1{i q_#1 R #1}
\def\unm{{1\over 2}}
\def\dmu#1{\mu{d #1\over d\mu}}
\def\bettyr#1{\beta_#1 := \dmu{#1^2_R}}
\def\boopr{\unm\gamma_{+R}\gamma_{-R}}

\def\lp{{\lambda^{(+)}}}
\def\lm{{\lambda^{(-)}}}
\def\lpm{{\lambda^{(\pm)}}}

\def\lsum{{\lp^2_R+\lm^2_R\over 2}}
\def\lmbb{\widehat{\lm}}
\def\lpbb{\widehat{\lp}}
\def\duc{2+(1-k)\kappa}

%\TagsOnRight
\nopagenumbers
\line{\hfil }
\rightline{ROM2F-92/60}
\vglue 1cm
\hsize=6.0truein
\vsize=8.5truein
\parindent=3pc
\baselineskip=10pt
\centerline{\elevenbf Quantum Models of Black Hole Evaporation }
\vglue 1.0cm
\centerline{\tenrm F.Belgiorno, A.S.Cattaneo }
\baselineskip=13pt
\centerline{\tenit Dipartimento di Fisica, Universit\'a di Milano and INFN sez.
di Milano}
\baselineskip=12pt
\centerline{\tenit 20133 Milano, Italy}
\vglue 0.3cm
\centerline{\tenrm F.Fucito }
\baselineskip=13pt
\centerline{\tenit Dipartimento di Fisica, Universit\'a di Roma II ''Tor
Vergata" and INFN sez. di Roma II}
\baselineskip=12pt
\centerline{\tenit 00173 Roma, Italy}
\vglue 0.3cm
\centerline{\tenrm and}
\vglue 0.3cm
\centerline{\tenrm M.Martellini}
\centerline{\tenit Dipartimento di Fisica, Universit\'a di Roma I
''La Sapienza"}
\baselineskip=12pt
\centerline{\tenit 00185 Roma, Italy}
\vglue 0.8cm
\centerline{\tenrm ABSTRACT}
The discovery of black-hole evaporation represented in many respects
a revolutionary event in scientific world; as such, in giving answers to
open questions, it gave rise to new problems part of which are
still not resolved.\par
\noindent Here we want to make a brief review of such problems and
examine some possible solutions.
\footnote{}{Invited Talk at the "Workshop on String Theory, Quantum Gravity
and the Unification of the Fundamental Interactions" Rome, September 21-26}
\vglue 0.3cm
{\rightskip=3pc
 \leftskip=3pc
 \tenrm\baselineskip=12pt%\parindent=1pc
 \noindent
\vglue 0.8cm }
\line{\bf 1. BH Evaporation: Open Questions \hfil}
The great problem of bh-evaporation is well exemplified
by the following question: which is the black-hole final state?\par
\noindent A complete answer to this question contains the solution to the
other great enigma: is there a loss of information in bh-evaporation?\par
\noindent That is, is it possible to recover the information contained in
the initial state of bh-evolution? In dimension $D=4$ these question have no
answer.
Hawking's work$^{11}$ is based on two distinct approximations:\par
a)the semiclassical approximation in which the bh background geometry is
classical,

b)the emitted radiation has no backreaction on the geometry.

It is well known that both these approximations fail at $m_{bh}\sim
M_{Planck}$;
at such energy scales it is necessary to make a full quantum treatement of the
gravitational field.

The so called ``Information Loss
Problem" consists in what follows: it is reasonable to think that we can
associate to black-holes a great entropy content, because we are not able to
measure and specify its internal microstates; our only physical possibility
to speak about a bh is to assign to it a macrostatus defined generally by three
measurable quantities: its mass, its angular momentum, its charge.\par
\noindent As soon as the bh forms, because of the event horizon, all the
informations
about the initial state and microscopic structure of the collapsing star
are lost. This is why we say that the ``bh has no hair".\par
The bh entropy is proportional to the area of its horizon. During the
evaporation the horizon shrinks causing the bh entropy to decrease. Can this
cause a loss of information?

This same problem can also be reformulated looking for the existence of
a unitary S-matrix$^{12}$.
\vglue 0.6cm
\line{\bf 2.  Bidimensional Models \hfil}
\vglue 0.4cm

Given the difficulties of the four dimensional case much work has been recently
concentrated on the study of bidimensional quantum models
that could give some insight, at least qualitatively, about the physical case.

Our starting point is the action containing the coupling of gravity to a
dilaton field
$$
S_{DG}={1\over 2\pi} \int d^2 x
\sqrt{-g} e^{-2 \Phi} (R+4 (\nabla \Phi)^2
+4 \lambda^2).
\eqno(2.1)
$$

As it is well known it is the coupling of the dilaton field $\Phi$ to the
Einstein metric that makes the theory non trivial.
The main advantages of such models are:

i) the theory is perturbatively renormalizable so that, at least in principle,
we can indagate its quantum aspects;\par
ii) we can think about $S_{DG}$ as deriving from an off-critical string theory
in a low energy approximation;\par
iii)after some manipulations the dilaton-gravity can be mapped into some
soluble model

The theory, using light-cone coordinates and in the conformal gauge
$g_{+-}=-{e^{2 \rho} \over 2}$ ($\rho$ is the Liouville mode) looks
like:

$$
S_{DG}={1\over \pi} \int d^2 \sigma [e^{-2 \Phi}(2 \partial_{+}\partial_{-}\rho
-4 \partial_{+}\Phi \partial_{-}\Phi+ \lambda^2e^{2 \rho})]
\eqno(2.2)
$$

\vglue 0.6cm
\line{\bf 3.  The CGHS Model \hfil}
\vglue 0.4cm

The purpose of this model$^6$ is to provide a description of:

i)the creation of a bh by means of matter shock waves

ii)the backreaction problem

This is achieved by summing to the action Eq.(2.1) a matter term:
$$
S_{M}={-1\over 4\pi} \int d^2 x g^{1\over 2} {\sum_{i=1}}^N (\nabla f_i)^2,
\eqno(3.1)
$$

where the $f_i$'s are conformal massless matter fields minimally coupled
to gravity whose energy tensor in light cone coordinates is proportional to
$\delta(x^+-x^0)$.

One way to keep the backreaction into account is to include at the Lagrangian
level the integrated trace anomaly which is local in our gauge:
the asymptotically flat coordinates trace anomaly contribution is:

$$
S_{anomaly}=-{N \over {12\pi}} \int d^2 \sigma \partial_{+}\rho
\partial_{-}\rho.
\eqno(3.2)
$$

In the large N limit, we can perform a 1/N expansion to find that the anomaly
term is of the same order of magnitude of the classical contribution.
Furthermore the quantum effects of gravity and of the dilaton can be neglected
in this approximation.

An analysis of the model reveals two different regions of the Penrose
diagram. The first is a weak coupling region where
$Exp(-2\Phi)\gg {N\over 12}$. The second is a strong coupling region where
$Exp(-2\Phi)\ll {N\over 12}$. These two regions are separated by a critical
time-like line $Exp(-2\Phi_{cr})={N\over 12}$.

The problem$^{5,13,14}$ is that $\Phi\sim\Phi_{cr}$ leads to a naked curvature
singularity or, in other words, the semiclassical approximation breaks down.
Moreover we find that the Hawking temperature is mass independent contrary to
our physical expectations.

\vglue 0.6cm
\line{\bf 4.  The Bilal-Callan-De Alwis Model \hfil}
\vglue 0.4cm
One possible way out of these problems$^{4,7,8,9}$ is to build  two-dimensional
models derived from low energy string theories using the many possible vacua
that lead to vanishing beta functions. The recipe to do this is:

i) to replace the ${N\over 12}$ coefficient in Eq.(3.2) with a generic
trace anomaly coefficient $\kappa$
to be determined by imposing the vanishing of the
total central charge;

ii) to modify the potential term by replacing it with conformal
vertices of weight (1,1) which in the classical limit
$\Phi \rightarrow -\infty$ reproduce the cosmological term.

The models thus obtained are Liouville-like. They are still ill-defined
at the quantum level as their kinetic energy goes to zero at a certain critical
value of the dilaton field for which also the curvature is singular.
A proposed way out of it is to restrict the fields to a region with no
singularity. These restriction are not natural from the point of quantum field
theory and up to now nobody has succeded in doing so consistently.

\vglue 0.6cm

\vglue 0.6cm
\line{\bf 5.  The CAT Model \hfil}
\vglue 0.4cm

We start with the following definitions:\par
\noindent
$\omega \equiv {Exp(-\Phi) \over {k^{1 \over 2}}}$\par
\noindent
$2 \Omega (\omega) \equiv \omega (\omega^2 -1)^{1 \over 2}-Log(\omega +
(\omega^2 -1)^{1 \over 2})$\par
\noindent
for $\omega \geq 1$ (as in ref. 4) and\par
\noindent
$2 \Omega^{\prime} (\omega) \equiv \omega (1-\omega^2)^{1 \over 2}-
ArcCos(\omega)$\par
\noindent
for $\omega \leq 1$\par
\noindent
$2 \chi \equiv \rho + (\omega)^2$.\par
\noindent It should be noted that the value $\omega = 1$ corresponds to the
critical value of the dilaton field.\par
The new idea is now the following: we shall formulate an effective theory with
the explicit aim of describing physics around $\omega = 1$ by mean of
a {\it non-local field redefinition}:\par
$2 \eta (\sigma) \equiv \Omega (\sigma + \epsilon)+
\Omega^{\prime} (\sigma - \epsilon)$\par
$2 \xi (\sigma) \equiv \Omega (\sigma + \epsilon)-
\Omega^{\prime} (\sigma - \epsilon)$\par
\noindent where $\sigma + \epsilon$, $\sigma - \epsilon$ are two nearby
points and $\omega (\sigma + \epsilon) \sim 1^{+}$,
$\omega (\sigma - \epsilon) \sim 1^{-}$.\par
\noindent The signature of the kinetic term thus obtained does not oscillate.

Now we have to construct interaction vertices in these new field variables
with conformal weight $(1,1)$
and match them against the cosmological constant term in the classical
limit.\par
\noindent

The resulting action is\par

$$
S_{cat}={1\over 2\pi} \int d^2 x [-k (\partial \chi)^2 +
(\partial \eta)(\partial \xi)+ \gamma_{+} e^{A_{+} \chi} +
\gamma_{-} e^{-A_{-} \chi+B \eta}]
\eqno(5.1)
$$

\noindent where the parameters appearing in the action are adjusted by the
request
of conformal invariance$^1$.\par
\noindent $S_{cat}$ describes an affine $\widehat{Sl(2)}$ Toda theory,
 previously
 studied by Babelon-Bonora in the framework of integrable models$^3$.\par
\noindent This is more evident if we make another field redefinition: $
\varphi \equiv i \sqrt{2\kappa} \chi.
$
We shall refer to our model
 as the conformal affine Toda black hole (CATBH) model.
 The CATBH model allows a standard perturbative quantization with the aim of
 unveiling the quantum effects of the dilaton-graviton fields: as a matter of
 facts, CAT fields are suitable functions of the
 dilaton and of the Liouville fields.\par

We now want to consider the renormalization flow of the classical
Babelon-Bonora action
$$
S_{BB}=\intx\left[{{1\over 2}\kinx{\varphi}+\de_\mu\eta\de^\mu\xi
-2\left({e^{2\varphi}+e^{2\eta-2\varphi}}\right) }\right].
\eqno(5.2)
$$
At the quantum level one must implement wave and vertex function
renormalizations so that in (5.2) one must introduce different
bare coupling constants in front of the fields as well as in front of
the vertex interaction terms. As a consequence one ends with the form
(5.1). However, according to the general spirit of the
renormalization procedure,
all generally covariant dimension 2 counterterms are possible in (5.1)
and hence also Feigin--Fuchs terms, i.e. the ones involving the $2D$-scalar
curvature, are allowed.
This ansatz is in agreement with the perturbative theory as
one could show following Distler and Kawai.
In our context the Feigin--Fuchs terms come naturally out if we look at
the improved stress tensor .
This leads us to consider the following quantum action:
$$
\eqalign {
S &=\intx\sqrt{ g}\left[ {
g^{\mu\nu} \left({{1\over 2}\de_\mu\varphi\de_\nu\varphi +
\de_\mu\eta\de_\nu\xi}\right)
+\gamma_+e^{i\lp\varphi}+\gamma_-e^{i\delta\eta-i\lm\varphi}
}\right.\cr
&+\curv\varphi +\curv\eta +\curv\xi ] .\cr }
\eqno (5.3)
$$
\par
Let us now start with the renormalization procedure of
(5.3) in a perturbative framework, in the hypothesis that
$\lm^2\sim\lp^2\sim4$.
Notice that $\xi$ plays the role of an auxiliary field; a variation with
respect to $\xi$ gives the on-shell equation of motion
$$
\nabla_\mu\nabla^\mu\eta=iq_\xi R,
\eqno(5.4)
$$
which in our perturbative scheme must be linearized around the
flat space:
$$
\de_\mu\de^\mu\eta=0.
\eqno(5.5)
$$
Following Ref.16, we define the renormalized quantities
at an arbitrary mass scale $\mu$ by:
$$
\matrix{
\eqalign{
\varphi &= Z^\unm_\varphi \varphi_R,\cr
\eta &= Z^\unm_\eta \eta_R,\cr
\xi &= Z^{-\unm}_\eta \xi_R,\cr}
&\eqalign{
\gamma_\pm &= \mu^2 Z_{\gamma_\pm} \gamma_{\pm R},\cr
\lpm^2 &= Z^{-1}_\varphi \lpm^2_R,\cr
\delta^2 &= Z^{-1}_\eta \delta^2_R,\cr}
&\eqalign{
 q^2_\varphi &= Z^{-1}_\varphi q^2_{\varphi R},\cr
 q^2_\eta &= Z^{-1}_\eta q^2_{\eta R},\cr
 q^2_\xi &= Z_\eta q^2_{\xi R}.\cr }
\cr}
\eqno(5.6)
$$
The following quantities are conserved through renormalization:
$$
\matrix {
r={q_\varphi\over\lm},
& l={\lp\over\lm},
& k=q_\xi\delta,
& p=q_\xi q_\eta.\cr }
\eqno(5.7)
$$
Following essentially the same procedures of Ref.16, with slightly
modifications due to the presence of extra fields and
by taking $\lp\not=\lm$, we
find that the theory can be renormalized at one loop
if we restrict ourself to the
on-shell renormalization scheme, i.e. if we get rid of the terms in
$\de_\mu\de^\mu\eta$ ( produced by the renormalization) using
(5.5). The true on-shell theory
should rely on (5.4), but at this perturbative order curvature terms
can be neglected: they are important only in the
renormalization of $\gamma_\pm$. The curvature terms are taken into
account$^{17}$
considering the modifications to the trace of the stress-energy tensor,
calculated on-shell. We finally get the following $\beta$ functions:
$$
\eqalign {
\beta_+:=\dmu{\gamma_{+R}}
&= 2\gamma_{+R}\left({{\lp^2_R\over 4}-1-q_{\varphi R}\lp_R}\right),\cr
\beta_-:=\dmu{\gamma_{-R}}
&= 2\gamma_{-R}\left({{\lm^2_R\over 4}-1+q_{\varphi R}\lm_R-q_{\xi R}
\delta_R}\right),\cr}
\eqno(5.8)
$$
$$
\eqalign{
\bettyr{\lpm} &= \boopr{\lp^2_R+\lm^2_R\over2}\lpm^2_R,\cr
\bettyr\delta &= \boopr\delta^4_R,\cr}
\eqno(5.9)
$$
$$
\eqalign{
\bettyr{{q_\varphi}} &= \boopr\lsum q^2_{\varphi R},\cr
\bettyr{{q_\eta}} &= \boopr\delta^2_R q^2_{\eta R},\cr
\bettyr{{q_\xi}} &= -\boopr\delta^2_R q^2_{\xi R}.\cr }
\eqno(5.10)
$$
Putting together the equations (5.9), we find that the following
quantity is also conserved through renormalization.\par
$$
d={1+l^2\over 4}{1\over\delta^2}-{1\over\lambda^2}.
\eqno(5.11)
$$
\par
This result
 needs not to be valid beyond this perturbative order.
Using the non-perturbative RG-invariants in (5.7) and dropping out
the rather cumbersome $R$ indices, we can rewrite the relevant
$\beta$ functions as:
$$
\eqalign {
\beta_+ &=2\gamma_+\left({
{l^2-4lr\over 4}\lm^2-1}\right)\cr
\beta_- &=2\gamma_-\left({
{1+4r\over 4}\lm^2-1-k}\right)\cr
\beta_\lm &= {1+l^2\over 4}\gamma_+\gamma_-\lm^4,\cr}
\eqno(5.12)
$$
Then we have to request that at a certain scale $t_0$, $t \equiv
log(\mu)$, both vertex operators have a conformal weight
$(1,1)$. As a consequence,
$\gamma_\pm$ depend on the scale $t-t_0$ only through
the cosh, which is an even function, and hence does not distinguish
between IR and UV scales. The requirement of having vertex
operators with the right conformal weight at the
renormalization point $t_0$
forces the theory to be dual (under the exchange
of the IR and UV scales). Explicitely we get:
\par
$$
\eqalign{
\lm^2(t) &= {8\over2+\kappa}\left\{{
1+k-{\duc\over\kappa}\tanh\left[{
{\duc\over\kappa}{2+k\over1+k}(t-t_0)}\right]}\right\},\cr
\gamma_-(t) &= A \left\{{
\cosh\left[{
{\duc\over\kappa}{2+k\over1+k}(t-t_0)}\right]}\right\}^{-2{1+k\over2+k}}.\cr}
\eqno(5.13)
$$
This UV-IR
duality implies that $\gamma_-(t)$ behaves in the same manner both at
an UV and at an IR scale with the asymptotic
behaviour:
$$
\matrix{
\gamma_-(t){\buildrel |t-t_0|\to\infty \over\sim}
A e^{-2s|t-t_0|},\hfil
&\hbox{ with }
&\hfil s={2+(1-k)\kappa \over\kappa}.\cr}
\eqno(5.14)
$$
where $\kappa=(N-23)/12$ and s is always positive and close to 1.\par
The main result here is that the RG-analysis$^1$ shows that
there exists an energy scale $t_{BC}\ll t_0$ at which our action ``becomes"
the de Alwis-Bilal-Callan action, in the sense that $
\left|{{\gamma_-(t_{BC})\over\gamma_+(t_{BC})}}\right| >> 1.
$
Notice that in the classical limit $e^\Phi\to 0$; our
effective action in the IR-phase $t=O(t_{BC})$ becomes the one of
ref. 4 since $\xi \to \eta \to \Omega$.

\par

Our strategy at this point is to use the above running coupling constants
$\gamma_-, \lambda_-$ to define ``effective" bh thermodynamic
quantities at the scale $t_{BC}$. Indeed by a
physical point of view the back-reaction should modify Hawking radiation
 emission and cause it to stop as soon as the bh has
radiated away all its initial
ADM mass,
 so that it should be reasonable to get a decreasing Hawking temperature
rate at the end point of bh-evaporation. In our CATBH model the
bh temperature is
proportional to $\mu \gamma_{-}^
{1 \over 2}$. We can regard $\gamma_{-}$ as a running coupling constant
in terms of $\mu$, which is roughly a measure of ADM mass.
In particular,
a reasonable ansatz for the bh solution
formed by N infalling matter shock waves,
allows us to find a simple
link between the RG-scale t and the bh mass (that is the
physical scale of our problem!). In fact a relation between the v. e. v.
of the operator-valued
scalar curvature $\sqrt{-g} R$ and the other CATBH running coupling constant
$\lambda_{-}$ can be obtained in the classical limit $e^\Phi\to 0$.
If we consider the
conformal gauge $\widehat{g_{\mu\nu}}=-{1 \over 2}
 e^{2\lambda\rho}\eta_{\mu\nu}$,
we get in the tree approximation:
$$
<\sqrt{-\hat g}\hat R>_{tree} = 2\lambda\partial_{\mu}\partial^{\mu}
<\rho>_{tree}.
\eqno(5.15)
$$
\par
\noindent
In eq. (5.15) we
may use the explicit solution describing the black
hole formation by $N$-shock waves $f_i$, with $<\rho>_{tree}$ given by the
CGHS classical solution:\par
$$
\eqalign{
e^{-2\lambda_{-}(t)<\rho(x^+)>_{tree}} &=
1-2\lambda_{-}(t)<\rho(x^+)>_{tree}+O(\lambda_{-}^2)\cr
&=-\kappa a(x^+-x_0^+)\theta(x^+-x_0^+)-\gamma_-(t)x^+x^-,\cr}
\eqno(5.16)
$$
where $\theta$ is the Heaviside function,
$x^\pm\equiv x^0 \pm x^1$ and $a\equiv{\rm const}$.
Therefore, at $x^+=x^+_0$, where the $f$-waves are sitting,
we have, using the light-cone
coordinates $x^\pm$:$
<\sqrt{-\hat g(x^+)}\hat R(x^+)>_{tree}\sim\gamma_-(t),
x^+\to x_0^+$.\par
\noindent

Since here we have two scales $x^+_0$ and $\mu$,
it is reasonable to set (in $c=\hbar=1$) $x^+_0\equiv{1\over\mu}$, since
the natural scale which describes the black hole formation is $x_0^+$.
Therefore we arrive to:
$$
<[\sqrt{-\hat {g}}{\hat R}](x_0^{+})>\sim\gamma_-[-log(x_0^{+})].
\eqno(5.17)
$$
Furthermore in the CGHS solution, the mass $m_{bh}$ of the bh created
grows linearly
with $x_0^+$, i.e. $m_{bh}\propto M^2_{\rm Pl}x_0^+$. Thus putting all
together we get that the end point of the bh-evaporation is charaterized
by:\par

$$
<\sqrt{-\hat {g}}{\hat R}>_{tree} \rightarrow 0, \hfill
m_{bh} \rightarrow 0,
\eqno(5.18)
$$\par

$$
T_{bh} \sim T_{0} ({m_{bh} \over m_{0}})^{s-1} \rightarrow 0,
\hfill
m_{bh} \rightarrow 0.
\eqno(5.19)
$$\par
\noindent
Notice that for ``astrophysical" bh, i.e. with large mass, one finds that\par

$$
T_{bh} \sim T_{0} ({m_{0} \over m_{bh}})^{s+1} \rightarrow 0,
\hfill
m_{bh} \rightarrow \infty.
\eqno(5.20)
$$

\noindent The vanishing of the bh temperature both for small and large bh
 mass is a consequence of the duality between UV and IR scales observed
above.\par
\noindent Then we see that the end-point of the bh-evaporation is
charaterized by a regular geometry and an almost zero
Hawking temperature.

There are various lines of developement of this model$^2$; in particular
we signal:
i)the calculation of the geometry and dilaton associated to the solutions
of our conformal affine Toda model; ii) the
 derivation of an (unitary?) exact S-matrix which is allowed (in principle)
by the quantum integrability of (5.3).

\vfill

\eject

\centerline {\bf References}\par

\item {1.}
F.Belgiorno,
A.S. Cattaneo, M. Martellini and F. Fucito, {\it A Conformal Affine
Toda Model of 2D Black Holes: a Quantum Study of the Evaporation End-Point }
to appear in Phys.Lett.B.
\item {2.}
F. Belgiorno,
A.S. Cattaneo, M. Martellini and F. Fucito, {\it A Conformal Affine
Toda Model of Two Dimensional Black Holes: the End Point State and the
S-Matrix}, work in progress.
\item {3.}
O. Babelon and L. Bonora , {\it Phys. Lett.} {\bf B244}(90)220.
\item {4.}
A. Bilal and C. Callan, {\it Liouville Models of Black Hole
Evaporation}, Princeton preprint PUPT-1320, hept@xxx/9205089, May 1992.
\item {5.}
B. Birnir, S.B. Giddings, J.A. Harvey and A. Strominger,
{\it Quantum Black Holes}, Santa Barbara preprint UCSB-TH-92-08,
hepth@xxx/9203042, \break March 1992.
\item {6.}
C.G. Callan, S.B. Giddings, J.A. Harvey, and A. Strominger
, {\it Phys. Rev.} {\bf D45}(92)1005.
\item {7.}
S.P. de Alwis, {\it Quantization of a Theory of 2d Dilaton Gravity
}, Colorado preprint COLO-HEP-280, hepth@xxx/9205069,
May 1992.
\item {8.}
S.P. de Alwis, {\it Black Hole Physics from
Liouville Theory}, Colorado preprint COLO-HEP-284, hepth@xxx/9206020,
June 1992.
\item {9.}
S.P. de Alwis, {\it Quantum Black Holes in Two Dimensions},
Colorado preprint COLO-HEP-288, hepth@xxx/9207095,
Jul 1992.
\item {10.}
S.B. Giddings and
A. Strominger, {\it Quantum Theory of Dilaton Gravity},\par
 UC Santa Barbara
preprint, UCSB-TH-92-28, hepth@xxx/9207034, July 1992.
\item {11.}
S.W. Hawking, {\it Comm.Math.Phys.} {\bf 43}(75)199.
\item {12.}
S.W. Hawking, {\it Phys. Rev.} {\bf D14}(76)2460.
\item {13.}
S.W. Hawking, {\it Evaporation of Two Dimensional Black Holes},
CALTECH preprint CALT-68-1774, March 1992.
\item {14.}
J.G. Russo, L. Susskind, and L. Thorlacius,
{\it Black Hole Evaporation in $1+1$ Dimensions},
Stanford preprint SU-ITP-92-4, January 1992;\par
L. Susskind and L. Thorlacius,
{\it Hawking Radiation and Back-Reaction},
Stanford preprint SU-ITP-92-12, hepth@xxx/9203054, March 1992.
\item {15.}
G. t'Hooft , {\it Nucl. Phys.} {\bf B335}(90)138.
\item {16.}
M.T.Grisaru, A.Lerda, S.Penati and D.Zanon
{\it Nucl. Phys.} {\bf B342}(90)564.
\item {17.}
A.B.Zamolodchikov
{\it JEPT Lett.} {\bf 43}(86)730.
\bye
\end
\bye